\documentclass[prl,floatfix,twocolumn,amsmath]{revtex4}
\usepackage{graphicx}
\usepackage{color}

\begin{document}

\def\ket#1{|#1\rangle}
\def\bra#1{\langle#1|}
\def\av#1{\langle#1\rangle}
\def\myarrow{\mathop{\longrightarrow}}
\def\ua{\uparrow}
\def\da{\downarrow}
\setlength\abovedisplayskip{9pt}
\setlength\belowdisplayskip{9pt}
\setlength\belowcaptionskip{-8pt}

\title{Strong coupling of an optomechanical system to an anomalously dispersive atomic medium}

\author{Haibin Wu$^{1}$ and Min Xiao$^{2,3}$ }
\affiliation{$^{1}$ State Key Laboratory of Precision Spectroscopy, Department of Physics, East China Normal University, Shanghai 200062, China}
\affiliation{$^{2}$ Department of Physics, University of Arkansas, Fayetteville, AR 72701, USA}
\affiliation{$^{3}$ National Laboratory of Solid State Microstructures and School of Physics, Nanjing University, Nanjing 210093, China}

\date{\today}

\begin{abstract}
We investigate a hybrid optomechanical system in which  a membrane oscillator is coupled to a collective spin of ground states of an intracavity $\Lambda$-type three-level atomic medium. The cavity field response is greatly modified by atomic coherence and the  anomalous dispersion generated by two Raman pumping beams near two-photon resonance. The optomechanical interaction, therefore radiation pressure force, is substantially enhanced due to superluminal propagation of photons in the cavity. Such improvement facilitates ground-state cooling of the mechanical oscillator with room temperature thermal environment. Moreover, it can greatly improve the sensitivity and bandwidth of  displacement measurement. In such system, optically-controlled strong-coupling interaction between the mechanical oscillator and cavity field could be implemented on small intracavity photon number, even at the single quanta level, which is important for weak-light nonlinear photonics  and the generation of nonclassical quantum states in the mechanical field.
\end{abstract}
\maketitle

Recent years have witnessed tremendous progresses on the cavity optomechanics, in which the mechanical degrees of the oscillator are coupled to cavity field via radiation pressure. Motivated by manipulating massive objects on quantum level~\cite{Weber}, developing more sensitive sensors~\cite{Caves}, providing quantum devices in future quantum state engineering, quantum information networking and quantum information processing~\cite{Lukin}, the controls over micro/nanomechanical oscillators have reached impressive levels. The sensitivity of displacement measurement has greatly enhanced with a precision down to the standard quantum limit~\cite{Simmonds,Heidmann,Abbott,Schwab,Kippenberg,Lehnert,Regal}. Sophisticated techniques for manipulating the mechanical modes have been developed and the ground-state cooling of mechanical modes has very recently been realized in a cryogenic pre-cooling setup~\cite{Connell,SimmodsGround,Painter,Painter2}. However, achieving ground-state cooling from a room temperature surrounding is still ongoing and very changeling~\cite{Kimble1}.

In seeking full quantum control of a mechanical oscillator, there are many proposals to realize hybrid optomechanical systems~\cite{Zoller, Hansch, Genes, Hammerer}, among which the coupling between atoms and the mechanical oscillator is seen to be a most promising way. This is mainly due to its large coherent time and well-controlled quantum toolbox at hand.  Linewidth narrowing by electromagnetically induced transparency (EIT) in a three-level atomic ensemble has been proposed to tailor the cavity response and facilitate cooling of the membrane~\cite{GenesEIT}. The coupling between an oscillator and long-distance atoms has been experimentally implemented in recent years~\cite{Kitching,Treutlein}.   However,  the dispersion properties of the atomic medium have not been really considered.  Moreover, in most experiments to date, optomechinical coupling is still quite weak and the strong interaction can only be realized in the high-finesse optical cavities and the classical, high-photon number regime.

In this Letter, we study an optomechanical system composed of a membrane and a dispersive $\Lambda$-type three-level atomic medium inside an optical cavity, as shown in Fig. 1. Two Raman pumping beams are used to generate a gain doublet for the probe (cavity) field which experiences  a small net gain and an anomalous dispersion~\cite{Wang}. We found that, in contrast to a normal dispersion which  essentially degrades the optomechanical coupling~\cite{GenesEIT},  this large anomalous dispersion  can greatly enhance the interaction between the oscillator mode and the cavity field. Its physical origin can be understood by considering the fact that a negative dispersion corresponds to superluminal propagation of photons in the cavity, which effectively increases the radiation pressure force. It can facilitate the ground-state cooling of the mechanical oscillator even at a room temperature thermal environment with a very low intracavity photon number avoiding the complex cryogenic pre-cooling. It can also greatly improve the sensitivity and bandwidth of the precision position measurement. The precision level below the zero-point fluctuation of the oscillator could be reached  on the few-quanta level. Such optically-controlled strong interaction can find many important applications in metrology, future quantum information processing, nonlinear photonics at the single-phonon level, and the generation of nonclassical quantum states in the mechanical field.

\begin{figure}[htb]
\includegraphics[width=3.2 in]{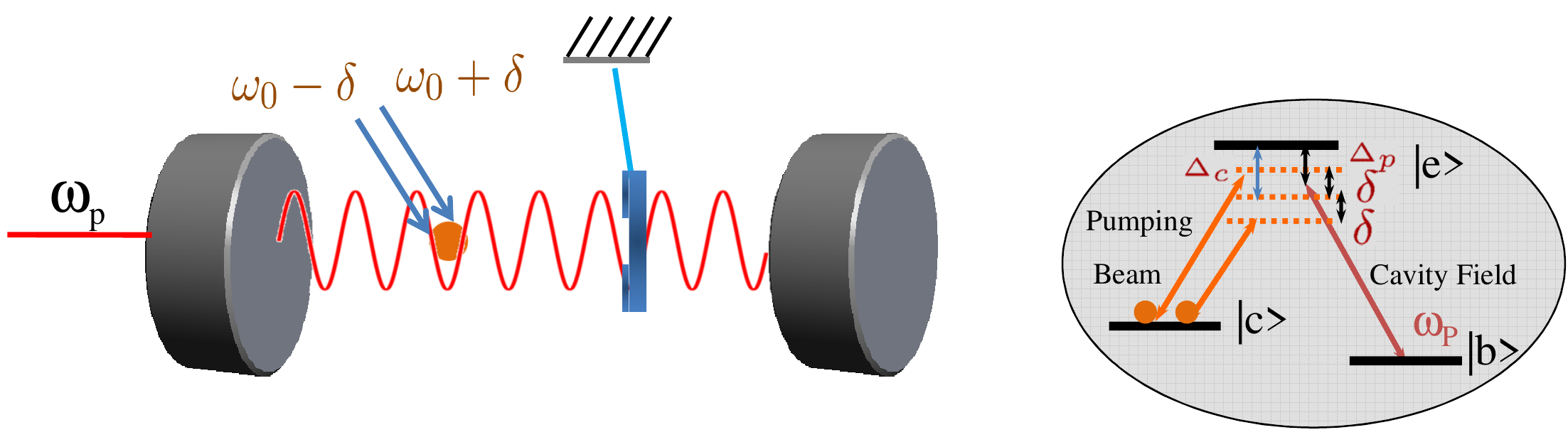}
\caption[example]
   { \label{fig:fig1}
Schematic diagram for a hybrid optomechanical system with a membrane and a dispersive three-level atomic medium in a high finesse optical cavity. Optical fields of frequencies $\omega_0+\delta$ and $\omega_0-\delta$ with Rabi frequencies $\Omega_{1}$ and $\Omega_{2}$, respectively, couple the ground state $\ket c$ to the excited state $\ket e$. The cavity field $a$ is driven by a laser with frequency $\omega_p$ and amplitude $a_{in}$. The cavity field couples the transition $\ket b$ to $\ket e$ with an interaction strength $g$.  }
\end{figure}
The basic system  is illustrated  in Fig. 1 with a membrane coupled to an intracavity $\Lambda$-type three-level atomic medium. Two optical fields (with frequencies $\omega_0+\delta$ and $\omega_0-\delta$, and amplitude ${\cal E}_1$ and ${\cal E}_2$, respectively) couple a metastable state $\ket c$ to an excited state $\ket e$. The cavity field $a$ is driven at the frequency $\omega_{p}$ and amplitude $a_{in}$. $\Delta_{p}=\omega_p-\omega_{eb}$ and $\Delta_{c}=\omega_0-\omega_{ec}$ are single-photon frequency detunings of the cavity field and pumping beams from atomic transitions $\ket c\rightarrow\ket e$ and $\ket b \rightarrow \ket e$, respectively. Two-photon detuning is defined as $\Delta_{pc}=\Delta_p-\Delta_c$. $\theta=\omega_{p}-\omega_{c}$ is the detuning of the driving field from the cavity resonant frequency $\omega_{c}$.   The membrane's vibrational frequency is $\omega_{m}$.
We write the Hamiltonian as
\begin{eqnarray}
H/\hbar&{=}&\omega_{c} a^{\dagger} a +\omega_m b^{\dagger}b+ \sum_{j=1}^{N}\sum_{i}^{b,e,c}\omega_i\ket i\bra i^j+H_{fa}  \nonumber\\
&+&ia_{in}(a^{\dagger} e^{-i\omega_{p}t}-ae^{i\omega_{p}t})+ G_{0}a^{\dagger}a(b+b^{\dagger}),
\label{eq:Hamiltonian}
\end{eqnarray}
where
\begin{eqnarray}
H_{fa}&=&-  \Omega\sum_{j=1}^{N} \ket {c}\bra {e}^{j} [e^{-i(\omega_{0}+\delta)t}+e^{-i(\omega_{0}-\delta)t}]\nonumber\\
&&-g \sum_{j=1}^{N}a\ket e\bra b^j+h.c . \nonumber
\end{eqnarray}
Here, $g$ is the atom-cavity coupling strength and $N$ the number of atoms in the cavity mode. $G_0$ is the optomechanical interaction strength. $\Omega=\bra e \mu_{ce}\ket {c}{\cal E}/\hbar$ is the Rabi frequency corresponding to the transition $\ket {c}\rightarrow\ket e$. For simplicity we have assumed ${\cal E}_1={\cal E}_2\equiv{\cal E}$.  Define $\sigma_{be}=\sum_{j=1}^N\sigma_{be}^{(j)}/\sqrt{N}$, $\sigma_{bc}=\sum_{j=1}^N\sigma_{bc}^{(j)}/\sqrt{N}$ for ${\cal E}_2=0$ and $\sigma_{bc}'=\sum_{j=1}^N\sigma_{bc}^{(j)}/\sqrt{N}$ for ${\cal E}_1=0$ as the collective atomic operators, where $\sigma_{be}^{(j)}$ and $\sigma_{bc}^{(j)}$ are the operators $\ket b\bra e^j$ and $\ket b\bra c^j$, respectively, for the $j$-th atom. We consider the typical Raman regime with the detunings $\Delta_p\approx\Delta_c\equiv\Delta_{0}$ to be much larger than $\Omega$ and all other decay rates, which allows us to adiabatically eliminate $\sigma_{be}$ and $\sigma_{ce}$. The atoms are initially prepared in the state $\ket c$. In the weak-cavity field limit, one can get the following set of coupled equations of motion.
\begin{align}
\dot a &=(i\theta-\kappa) a -iC (\sigma_{bc}+\sigma_{bc}')-i G_0 a (b+b^{\dagger})+\tilde{a}_{in},\nonumber\\
\dot\sigma_{bc} &= [i(\Delta_{pc}+\delta)-\gamma_{bc}]\sigma_{bc}+iCa+\tilde{\sigma}_{bc}^{in},\nonumber\\
\dot\sigma_{bc}' &=[i(\Delta_{pc}-\delta)-\gamma_{bc}]\sigma_{bc}'+iCa+\tilde{\sigma}_{bc}^{in},\nonumber\\
\dot b&=(-i\omega_m-\gamma_{m})b-iG_0a^{\dagger}a+\tilde{\xi},
\label{e2}
\end{align}
where $C=g\sqrt{N}\Omega/\Delta_0$ gives the effective coupling between the cavity field and the collective atomic spin $\sigma_{bc}$. The decay rates are $\kappa$ for the cavity field, $\gamma_{bc}$ for the atomic coherences, and $\gamma_{m}$ for the membrane, respectively. The cavity driving field is given by $\tilde{a}_{in}$; $\tilde{\sigma}_{bc}^{in}$ and $\tilde{\xi}$ are the quantum noise operators. Their correlations are $\langle\tilde{a}_{in}(t)\tilde{a}_{in}^{\dagger}(t')\rangle=2\kappa\delta(t-t')$, $\langle\tilde{\sigma}_{bc}^{in}(t)\tilde{\sigma}_{cb}^{in}(t'\rangle=2\gamma_{bc}\delta(t-t')$ and $\langle \tilde{\xi}(t)\tilde{\xi}^{\dagger}(t'\rangle=2\gamma_{m}\delta(t-t')$, respectively.

 For this dual-pump system, a gain doublet is generated with the susceptibility given by~\cite{Wang}
\begin{align}
\chi=i M\bigg(\frac{1}{i(\Delta_{pc}+\delta)-\gamma_{bc}}+\frac{1}{i(\Delta_{pc}-\delta)-\gamma_{bc}}\bigg),
\end{align}
where $M\equiv C^2/\omega_{c}$.  The real part (dispersion) and imaginary part (gain or loss) of the susceptibility are plotted in Fig. 2 (a). It is clearly shown that there is a large anomalous dispersion and a small residual gain near the line center. The cavity field in the steady state can be written as ~\cite{Clerk}
\begin{align}
a=g \tilde{a}_{in}+\sqrt{|g|^2-1}d^{\dagger}
\end{align}
where $g=[i\Delta-\kappa-C^2(\frac{1}{i(\Delta_{pc}+\delta)-\gamma_{bc}}+\frac{1}{i(\Delta_{pc}-\delta)-\gamma_{bc}})]^{-1}$ is the effective gain and $\Delta\equiv\theta-2 G_{0}|a|^2/\omega_{m}$ is the effective cavity field detuning including the radiation pressure-induced optical resonance shift, respectively. $d^\dagger$ is a bosonic operator representing the noise added by the amplifier, which is required in such systems to preserve the commutation relation for the cavity field operator $a$~\cite{Clerk, Boyd}.  The cavity transmission spectrum (Fig. 2(b)) shows a three-peak structure, which is analogous to the case observed in cavity EIT ~\cite{Wuprl}.
\begin{figure}[htb]
\includegraphics[width=3.2 in]{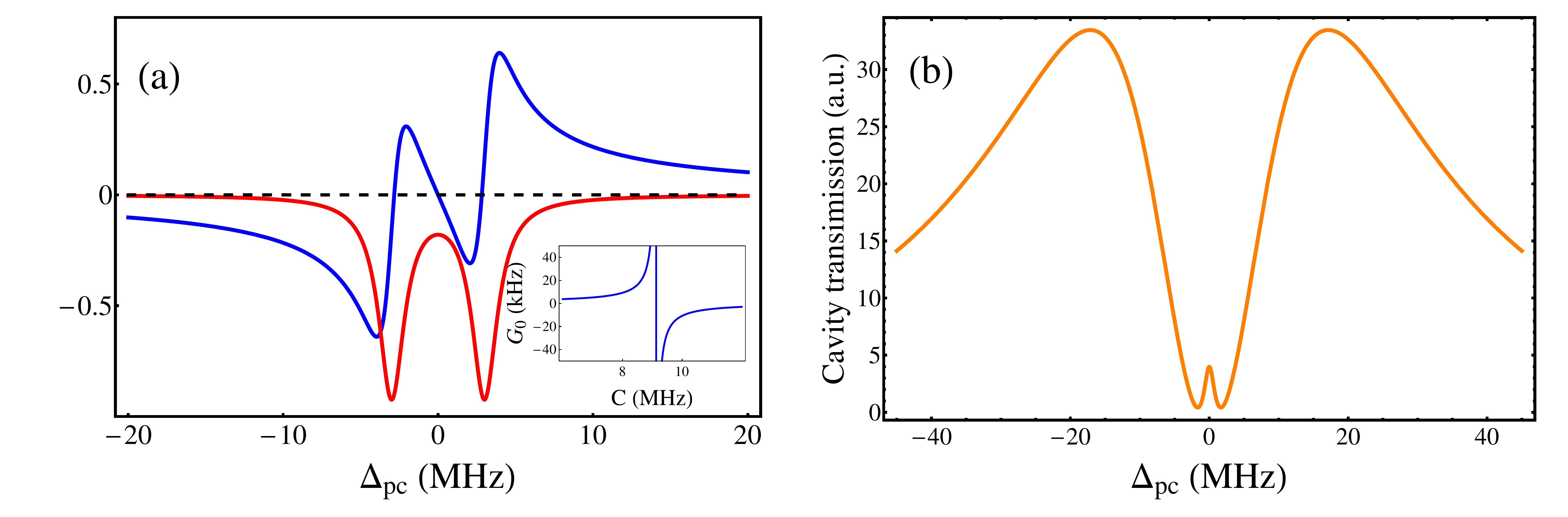}
\caption[example]
   { \label{fig:fig2}
(a) The dispersion ($ Re(\chi)$, blue curve) and gain ($Im(\chi)$, red curve) of atomic medium as a function of the two-photon detuning $\Delta_{pc}$. (b) The cavity transmission vs $\Delta_{pc}$. Inset: the interaction strength $G_0$ between the oscillator and cavity field as a function of Rabi frequency $\Omega$.  }
\end{figure}

With such dispersive medium in the cavity, the cavity response is greatly changed. The cavity resonant frequency is $\omega_c=j c\pi/(L+Re(\chi)l/2)$, where $j$ is an integer, $c$ the speed of the light in vacuum, $L$ the cavity length and $l$ the length of medium respectively. Radiation pressure force causes the instantaneous displacement $x$ of the membrane. A variation of length corresponds to a variation of the frequency $\delta \omega =-\eta_c \frac{\omega_c}{L}\frac{1}{1+\eta}x$, where $\eta=\frac{l\omega_c}{2L}\frac{\partial Re(\chi)}{\partial\omega}|_{\omega=\omega_c}$ and $\eta_{c}$ is a constant determined by the properties of the membrane. Correspondingly, the coupling strength of the optomechanical system can be written as
\begin{align}
G_{0}&=\eta_{c}\omega_{c}\frac{L_{m}}{L}\frac{1}{1+\eta},
\end{align}
where $L_{m}=\sqrt{\hbar/2m\omega_{m}}$ is the zero point motion of the mechanical oscillator mode ($m$ is the mass of the membrane). For the case of anomalous dispersion, $\eta<0$. $G_0$ can be optically tuned by  $\Omega$ and the frequency detunings. When $\eta$ gets close to $-1$, the coupling strength $G_0$ between the mechanical oscillator and cavity field is greatly enhanced, as shown in inset of Fig2. (a).

Linearizing Eq. 2 near a stable steady state solution by mapping $o\rightarrow \bar {o}+o(t)$, where $\av {o(t)}=0$, we get the fluctuation equations as
\begin{align}
\dot a &=(i\Delta-\kappa) a -iC (\sigma_{bc}+\sigma_{bc}')-i G(b+b^{\dagger})+\tilde{a}_{in},\nonumber\\
\dot\sigma_{bc} &= [i(\Delta_{pc}+\delta)-\gamma_{bc}]\sigma_{bc}+iCa+\tilde{\sigma}_{bc}^{in},\nonumber\\
\dot\sigma_{bc}' &=[i(\Delta_{pc}-\delta)-\gamma_{bc}]\sigma_{bc}'+iCa+\tilde{\sigma}_{bc}^{in},\nonumber\\
\dot b&=(-i\omega_m-\gamma_{m})b-iG (a+a^{\dagger})+\tilde{\xi},
\label{e2}
\end{align}
where $G=G_{0}\bar{a}\equiv G_0\sqrt{\langle n_c\rangle}$. Let the cavity detuning $\Delta$ and two-photon detuning $\Delta_{pc}$ match to the anti-Stokes motional frequency with $\Delta=-\omega_m$ and $\Delta_{pc}=\omega_m$, the final mechanical mode occupancy is given by
\begin{align}
n_f=\frac{\gamma_m}{\gamma_m+\gamma}n_i+ \frac{G^2\kappa+4G^2C^2\gamma_{bc}^2D_c/(\delta^2+\gamma_{bc}^2)^2}{(\gamma_m+\gamma)[4\omega_m^2+\kappa_{eff}^2/(1+\eta)^2]}.
\end{align}

$4G^2C^2\gamma_{bc}^2D_c/(\delta^2+\gamma_{bc}^2)^2$ is the contribution from  ground atomic  decoherence, which is typically very small. $\kappa_{eff}=\kappa-C^2 2\gamma_{bc}/(\delta^2+\gamma_{bc}^{2})$ is the effective cavity decay including the contribution of atomic coherence.  $\gamma$ is the effective cooling rate, $\gamma\approx G^2 (1+\eta)/\kappa_{eff}$. Here, the cavity linewidth broadening due to the negative dispersion is given by the term $1+\eta$ in the formula ~\cite{Haibin}. The first term is the thermal occupancy, which can be explicitly expressed as
\begin{equation}
n_{thermal}=\frac{\gamma_m}{\gamma_m+\eta_{c}^{2}L_{m}^{2}\omega_{c}^{2}\langle n_c\rangle/[L^2(1+\eta)\kappa_{eff}]}n_i.
\end{equation}

With a large anomalous dispersion generated by the dichromatic pumping, the interaction between light field and mechanical oscillator is enhanced by a factor of $1/(1+\eta)$, so the thermal noise is greatly suppressed. Moreover, with $C^2\rightarrow \kappa (\delta^2+\gamma_{bc}^{2})/(2\gamma_{bc})$, $\kappa_{eff}\rightarrow 0$. If $\kappa<\gamma_{bc}$, the atomic dispersion $\eta$ is small in the case. Such small $\kappa_{eff}$ caused by the atomic coherence effectively increases the cooling rate $\gamma$. This scheme is especially attractive for the optomechanical system with high initial temperature. The resulting cooling scheme does not require resolved-sideband condition for the cavity and the pre-cooling of the membrane, which relaxes the constraints in current standard optomechanical cooling schemes. It becomes possible to directly make ground-state cooling of the mechanical modes from room temperature.

When the dispersion is small and  cavity linewidth is  narrow compared to the resonant frequency of the mechanical modes, i.e., $\omega_{m}\gg \kappa$, in the so-called resolved-sideband regime, the mode occupancy from radiation pressure becomes $(\kappa/2\omega_m)^2$ at the strong coupling regime, which is consistent with the resolved-sideband optomechnical cooling limit ~\cite{Kippenberg2,Girvin2,GenesCoolinglimit}. However, with the strong dispersion, especially as $\eta$ gets close to $-1$, the cooling limit for radiation pressure is  given by $(1+\eta)\kappa/\kappa_{eff}$, which can be made very small compared to the typical cooling limit of $(\kappa/2\omega_m)^2$.
\begin{figure}[htb]
\includegraphics[width=2.5 in]{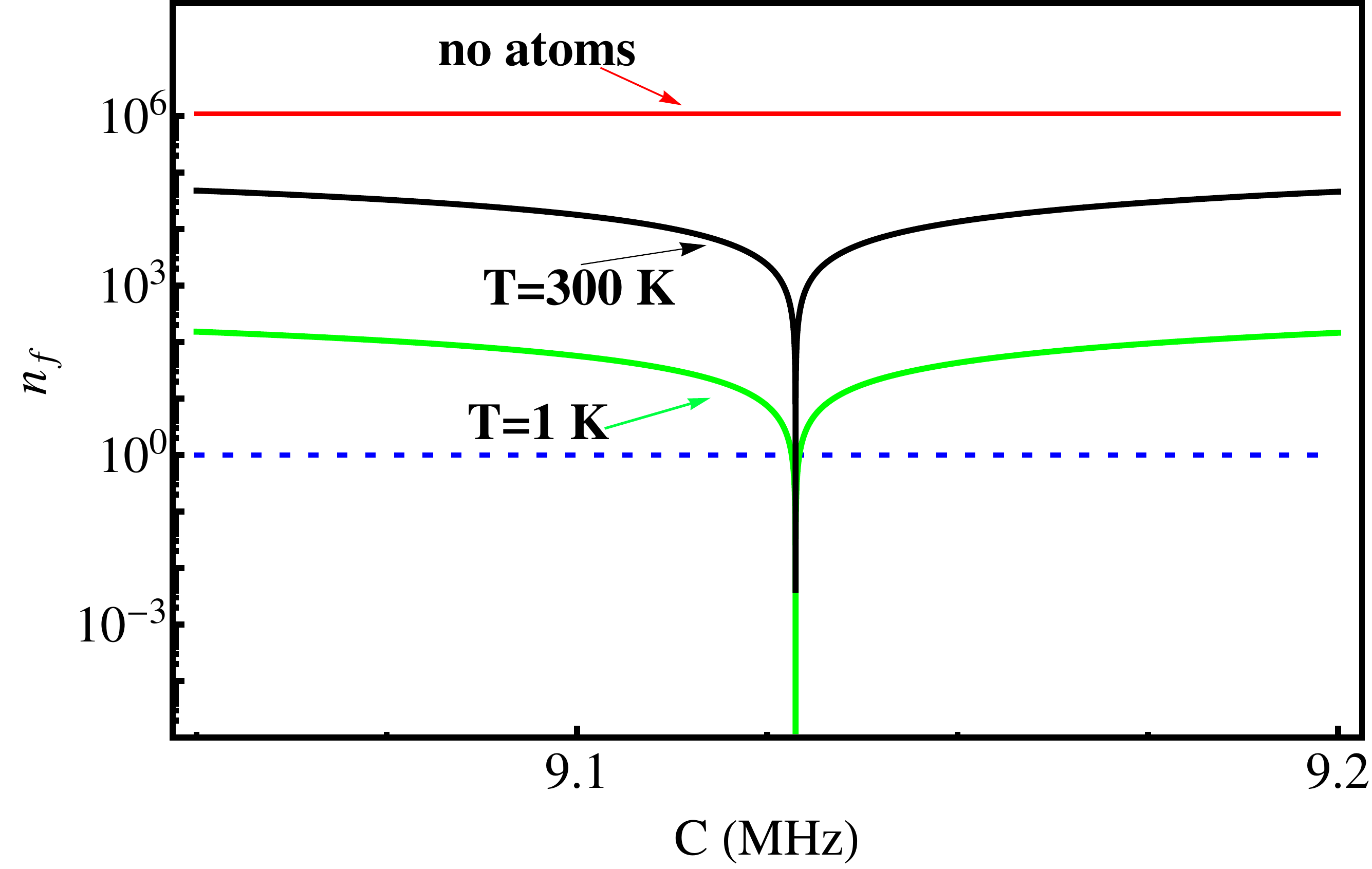}
\caption[example]
   { \label{fig:fig3}
Final mode occupancy $n_f$ as a function of $C$ with the average intracavity photon number $\langle n_c\rangle=10$ under the conditions of $\Delta=-\omega_m$ and $\Delta_{pc}=\omega_m$. Other parameters are: $\omega_m=2\pi\times 200\,$kHz, quality factor $Q_m\approx 10^7$, the effective mass 1 ng. The cavity length $L=1\,$ cm, the length of atomic medium $l=1\,$ mm, cavity decay rate $\kappa=2\pi \times 1\,$MHz. $\delta=2\pi\times 2\,$MHz and $\gamma_{bc}=2 \pi\times1\,$ MHz. The membrane temperature is initially at about 300 K with occupancy $10^7$ for red curve (no atoms) and black curve (counting the atomic dispersion). The green curve is initially at  $T=1 K$. The dashed (blue) line represents $n_f=1$. }
\end{figure}

To better illustrate such cooling effect, we now take a practical optomechanical system as an example. The membrane vibrational frequency is $\omega_{m}=2\pi\times 200\,$kHz, quality factor $Q_m\approx 10^7$ and an effective mass 1~$ng$ ~\cite{JHarris}. The thermal temperature is chosen to be 300 K (room temperature) with an initial occupancy $n_i=10^7$ or 1 K with initial occupancy $n_i=10^5$. The cavity is confocal with a length of 1 cm and a finesse of about $10^4$, corresponding to a cavity decay rate of $\kappa= 2\pi \times 1\,$MHz. The single-photon coupling strength without considering the dispersion is about $G_0=2\pi\times 300\,$Hz. The atomic medium is chosen to be a cold $^{87} Rb$ cloud with $10^9$ atoms in 1$mm^3$.  The detuning $\delta=2\pi\times2$ MHz and atomic decay rate $\gamma_{bc}=2\pi\times1\,$MHz are taken from Refs.~\cite{Wang, Pati}. The final mode occupancy $n_f$  as a function of Rabi frequency $\Omega$ is shown in Fig. 3 with an average intracavity photon number $\langle n_c \rangle=10$ when $\Delta=-\omega_m$ and $\Delta_{pc}=\omega_m$.  The red curve is for a membrane temperature of $T=300$ K with no atoms; black curve and green curve are for membrane temperatures of $T=300$ K and $T=1$ K, respectively, with taking large dispersion into account. It is clearly seen that without the atoms the cooling of the optomechanical oscillator is very weak (red curve) due to a small coupling strength. However, the cooling mediated by a negatively dispersive medium and atomic coherence becomes very efficient. The oscillator can be cooled from room temperature to its ground state in the case of weak light (small average intracavity photon number) (black curve in Fig. 3).

The strong cooling effect can be understood from the enhanced radiation pressure force. For such a hybrid optomechanical system, the force due to the momentum flips of  photons reflected from the membrane is given by
\begin{equation}
F=\frac{2\hbar k}{t_{r}}|a(t)|^2=\frac{\hbar \omega_c}{L-(1-c/v_g)l}|a(t)|^2 ,
\end{equation}
where $t_r$ stands for the round-trip time and $|a(t)|^2$ is the intracavity intensity normalized to photon number. $v_g=c/[1+\omega_c \partial Re(\chi)/(2\partial \omega)]$ is the group velocity of the cavity field in the atomic medium. The lifetime of photons in the cavity is increased for the case of normal dispersion. For such Raman system, the large anomalous dispersion near the center of gain doublets causes $v_g>c$ or even negative $v_g $, corresponding to the superluminal propagation of light. It shortens the round-trip time of intracavity photons and therefore effectively increases the radiation pressure force, which can greatly enhance cooling efficiency for the mechanical modes.

\begin{figure}[htb]
\includegraphics[width=2.3in]{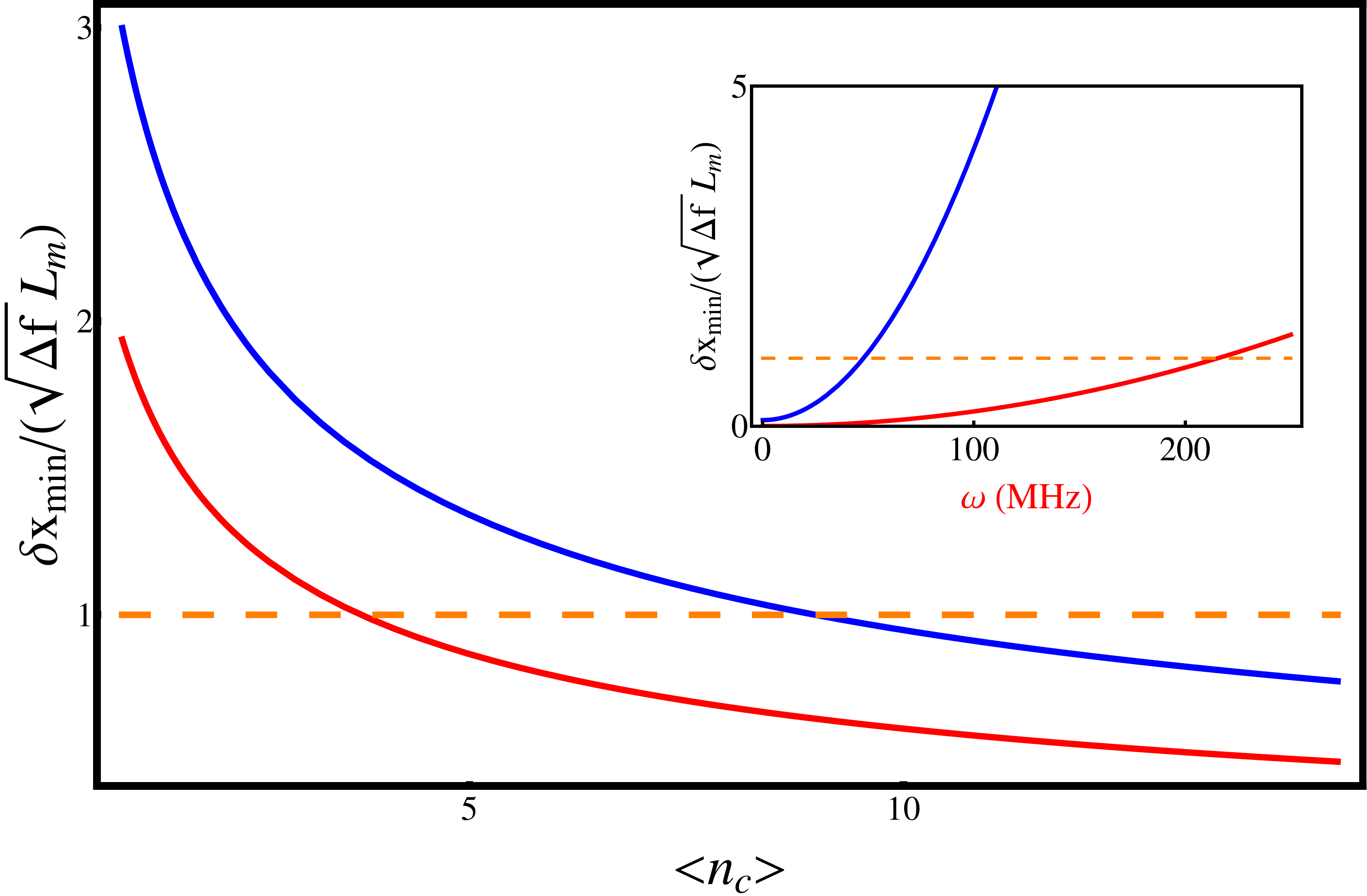}
\caption[example]
   { \label{fig:fig4}
$\delta x_{min}/(\sqrt{\Delta f}L_m)$ as  a function of the average intracavity photon number $\langle n_c\rangle$. The blue curve is for the case without atoms; the red curve is the case enhanced by the dispersion. Inset:  $\delta x_{min}/(\sqrt{\Delta f}L_m)$  as a function of the measurement frequency $\omega$, where $\langle n_c\rangle=10$ and $\kappa=2\pi \times15$ MHz. Other parameters are the same as Fig. 3.   }
\end{figure}

This scheme can also be used to greatly improve the sensitivity of displacement measurement. With the cavity input-output relation, $a_{out}=a_{in}-\sqrt{\kappa}a$, the symmetrical noise spectral density \cite{nsd} of the phase quadrature of the cavity output field can be derived. For a resonant probe beam, the minimum detectable position change $\delta x_{min}$ in a displacement measurement~\cite{Kippenberg} is calculated to  be
\begin{equation}
\frac{\delta x_{min}^2}{\Delta f}=\frac{\omega^2+[\kappa_{eff}/(1+\eta)]^{2}+4C^2\kappa\gamma_{bc}^{3}/(\delta^2+\gamma_{bc}^{2})^2}{4G_{0}^{2}\langle n_c\rangle \kappa/[L_{m}^{2} (1+\eta)^2]},
\label{eq:sensitivity}
\end{equation}
where $\Delta f$ is the measurement bandwidth and the background noise from the quantum fluctuation has been normalized to 1. $4C^2\kappa\gamma_{bc}^{3}/(\delta^2+\gamma_{bc}^{2})^2$ is spin noise of the atomic ground-state  which depends on the decay rate between atomic states $\ket b$ and $\ket c$. When neglecting this term, the minimum detectable displacement becomes the typical result except the enhanced interaction and modified cavity decay rate by atomic coherence. It is easily seen that the sensitivity and bandwidth of position measurement are substantially improved due to the large negative dispersion, as shown in Fig.~\ref{fig:fig4}. This scheme is different from the ones used in typical optomechanical systems in which the sensitivity is mainly increased by the cavity finesse and intracavity photon number. In this hybrid optomechanical system, the detectable displacement can reach an unprecedented sensitivity at small intracavity photon number.

When $-1<\eta<0$, the interaction between the cavity field and mechanical oscillator is greatly enhanced, which will benefit the ground-state cooling and sensitivity measurement. It would be especially interesting to  see what will happen when $\eta=-1$. At $\eta=-1$, it is under the condition of so-called ``white-light cavity" (WLC) ~\cite{Wicht, Pati,Haibin}, which was first proposed in the gravitational-wave detection scheme for increasing the buildup of intracavity field without sacrificing the detector bandwidthcavity linewidth.  It has been demonstrated that the dispersion variation by parameter deviation close to the point could be controlled within $10^{-4}$ ~\cite{Qing}, which makes the robustness of the scheme for such a hybrid atom-optomechanical system. It is clear in Eq.~\ref{eq:sensitivity} that the bandwidth of the displacement measurement can be greatly increased near WLC, which is very important because most optomechanical studies work only in the resolved sideband regime ($\omega_m\gg \kappa$) limiting the measurement bandwidth. Since we are working at the regime of small effective gain and large driving, the noise generated from the gain has little effect to the final occupancy of the mechanical oscillator and the precision position measurement. Although the possibility of the superluminal propagation of light on the one- or few-photon level has been studied previously~\cite{Stern}, the contribution of noise from the linear amplification process would become more and more important for a vacuum input and therefore cannot be neglected anymore. Another interesting result is that the quantum noise of vacuum field associated with an anomalously dispersive cavity can exceed $\hbar \omega/2$~\cite{Rosa}. How to understand the quantum noise limit for such optomechanical systems with anomalous dispersion remains to be an open question and needs further investigation.

In conclusion, a hybrid optomechanical system,  with a membrane as an oscillator and an intracavity dispersive atomic medium,  has been proposed and investigated. The interaction between  the collective atomic spin of ground states and the membrane is mediated by the cavity field. The large negative dispersion generated by the pumping fields greatly enhances the optomechanical coupling.  It makes ground-state cooling of phonon modes possible from a room temperature thermal environment. The sensitivity and bandwidth of displacement measurement can be substantially improved. This scheme paves a new way to reach the strong coupling regime of cavity quantum optomechanics, which can have many important applications in  future quantum state engineering, such as  nonlinear photonics at the single-phonon level and the generation of nonclassical quantum states in the mechanical fields.

We thank M. Fleischhauer for useful discussions. This work is partially supported by the National Natural Science Foundation of China under Grant No. 11374101 and Shanghai Pujiang Program under Grant No. 13PJ1402500.

\bibliographystyle{amsplain}

\begin{thebibliography}{32}
\bibitem[{nsd()}]{nsd}
\bibinfo{note}{The symmetrized noise spectral density is defined by
  $S_{xx}^{out}(\omega)=\frac{1}{2}\big(S_{x}^{out}(\omega)+S_{x}^{out}(-\omega)\big)$,
  where $2\pi\delta(\omega+\omega')S_{x}^{out}(\omega)=\big\langle
  x(\omega)x^{\dagger}(\omega')\big\rangle$ and $x(\omega)=-i
  \big(a_{out}(\omega)-a_{out}^{\dagger}(\omega)\big)$.}
\end{thebibliography}

\end{document}